\documentclass[11pt,a4paper]{article}
\usepackage[T1]{fontenc}
\usepackage[english]{babel}
\usepackage{lmodern}
\usepackage{amsmath,amssymb}
\usepackage{booktabs,tabularx,array}
\usepackage[margin=1in]{geometry}
\usepackage{enumitem}
\usepackage{xcolor}
\usepackage{url}
\usepackage{hyperref}
\hypersetup{
 colorlinks=true,linkcolor=blue!50!black,citecolor=blue!50!black,urlcolor=blue!50!black,
 pdftitle={MUB: An Identity-Masked Benchmark for Music Understanding in Multimodal Language Models},
 pdfauthor={Kun Liu; Lyra Tao},
 pdfsubject={Tempo, key, metre, and structural segmentation evaluation},
 pdfkeywords={music understanding, audio-language models, benchmark, structural segmentation}}
\setlist{leftmargin=*,itemsep=2pt,topsep=4pt}
\title{\textbf{MUB: An Identity-Masked Benchmark for Music Understanding in Multimodal Language Models}\thanks{Code and data: \url{https://github.com/Token-Spark/music-understanding-benchmark}}}
\author{Kun Liu\\\href{mailto:founder@tokenspark.uno}{\texttt{founder@tokenspark.uno}}
\and Lyra Tao\\\href{mailto:wenyumusic@qq.com}{\texttt{wenyumusic@qq.com}}}
\date{October 2026}

\begin{document}
\maketitle

\begin{abstract}
Evaluating musical perception in multimodal language models requires separating audio evidence from answers recoverable through metadata or external tools. We present the Music Understanding Benchmark (MUB), a locally assembled collection of 500 identity-masked recordings with existing annotations for tempo, key, time signature, and structural intervals. The collection covers all 24 major/minor key combinations, tempos of 47--180 beats per minute, and 4,676 annotated intervals. Original titles, artists, and platform identifiers are excluded from evaluation inputs. Relative to recent music question-answering and temporal-grounding work, MUB emphasizes scalar tempo prediction and full-recording interval outputs, together with a documented metadata-masking protocol. Saved baseline predictions yield 53.6\% tempo accuracy for a simple DSP estimator and 98.4\% time-signature accuracy for a constant 4/4 predictor. We identify important limitations, including an asymmetric legacy key metric, unverified independent annotation agreement, extreme metre imbalance, and unresolved recording-level rights clearance. No contemporary multimodal language model comparison has yet been completed. MUB is therefore an evaluation resource and protocol under validation, rather than a completed demonstration of model capabilities.
\end{abstract}

\section{Introduction}
\label{sec:intro}
Music understanding operates across interacting temporal and tonal scales. A listener may infer a pulse, organize beats into metre, identify a tonal centre, and locate recurring sections. Fluent verbal descriptions alone do not establish accuracy on these operations. Evaluation must also distinguish direct use of audio from answer-distribution priors, memorized recording associations, and external retrieval.

Recent audio-language systems have expanded considerably. Audio Flamingo 3 supports speech, sound, and music, including long-audio interaction~\cite{af3}; Music Flamingo focuses on richer musical understanding~\cite{music-flamingo}. Evaluation has also progressed beyond broad descriptions: MuChoMusic uses human-validated multiple-choice questions~\cite{muchomusic}, HumMusQA emphasizes human-written expert questions~\cite{hummusqa}, GaMMA combines global and temporal understanding~\cite{gamma}, and MusTBench targets temporal grounding~\cite{mustbench}. Coverage of tempo, key, metre, and structure is consequently not itself a new research contribution.

MUB brings together four outputs for each of 500 recordings: a scalar tempo, a tonic and mode, a time signature, and a sequence of structural intervals. It reuses an existing music-annotation corpus, masks explicit recording identities in the inference interface, and supplies simple baseline and scoring implementations. The contributions of this version are:
\begin{enumerate}
 \item A documented 500-recording selection with all 24 major/minor keys and 4,676 structural intervals.
 \item A metadata-masking protocol that removes original titles, artists, platform song identifiers, task identifiers, and annotator identities from model-facing inputs.
 \item An inspectable evaluation pipeline with waveform-based and answer-distribution reference predictions, together with an audit of its current limitations.
\end{enumerate}

These are resource and protocol contributions. Identity masking reduces an explicit title-based shortcut but cannot establish that a recording was absent from training. Baseline agreement is a consistency observation, not proof of label validity. The present study does not rank multimodal models or establish a general measure of musical understanding.

\section{Related Work}
\label{sec:related}
\subsection{Music Question Answering and Human Curation}
MuChoMusic~\cite{muchomusic} introduced 1,187 human-validated multiple-choice questions over 644 tracks and evaluated five audio-language models. Its analysis of reliance on linguistic information motivates controls against shortcuts. We do not assume that MuChoMusic exposed recording titles to evaluated models, nor characterize its human-validated labels as purely automatic. MUB differs primarily in its output format and reused corpus, not in being the first human-curated music benchmark.

HumMusQA~\cite{hummusqa}, first submitted in March 2026, provides 320 expert-written and validated questions over 108 tracks. Its coverage includes rhythm, harmony, and musical structure. It demonstrates that professional musical reasoning and human curation already form part of the evaluation landscape. MUB instead derives global labels and full-recording intervals from an existing annotation workflow; the independence and agreement of its annotators remain to be measured.

\subsection{Recent Audio-Language Models and Representation Access}
Audio Flamingo 3~\cite{af3}, first submitted in July 2025, advances unified audio-language modelling with long-audio support, multi-audio interaction, and optional reasoning. Music Flamingo~\cite{music-flamingo}, first submitted in November 2025, specializes in music understanding and uses musical concepts including tempo, key, metre, and structure in its training approach. These systems motivate task-level evaluation, but their existence should not be treated as evidence of their performance on MUB: neither has been evaluated here.

Fang et al.~\cite{gpt-judge}, first submitted in January 2025, investigate GPT as a judge for music-understanding annotations using symbolic representations and prompting, including beat, chord, and key information. This is relevant to annotation auditing, but representation-assisted judgement is different from inferring musical properties directly from a waveform. A native-audio track must disclose and constrain such intermediate representations rather than silently treating them as native perception.

\subsection{Global and Temporal Music Understanding}
GaMMA~\cite{gamma}, first submitted in May 2026, has substantial overlap with MUB. Its MusicBench contains 3,739 questions, comprising 2,741 global and 998 temporal questions, including BPM, key, metre-related rhythm, and section identification within supplied intervals. It also evaluates structural segmentation separately on Harmonix using segment matching. MUB therefore cannot claim first or unique coverage of the four task areas. Its more specific focus is the joint packaging of identity-masked recordings, scalar global predictions, and full-recording interval outputs from a particular annotation corpus.

MusTBench~\cite{mustbench}, first submitted in May 2026, provides 1,264 expert-validated questions over 517 songs for five temporal-grounding tasks. These include source timing, local transitions, ordering, and mood intervals. Such temporal questions complement, but do not coincide with, exhaustive segmentation of a recording into formal sections. Conversely, MUB's present label-agnostic interval metric does not establish the richer temporal reasoning tested by MusTBench.

Table~\ref{tab:related} summarizes the relevant distinctions. Dates refer to first preprint submission rather than later revisions; all six recent works discussed above fall within 2025--2026. These citations establish overlap and context, not the novelty or validity of MUB's labels.

\begin{table}[tbp]
\centering\small
\begin{tabularx}{\linewidth}{@{}p{0.22\linewidth}p{0.12\linewidth}X@{}}
\toprule
Work & Year & Evaluation or modelling emphasis \\
\midrule
MuChoMusic & 2024 & Human-validated music multiple-choice questions. \\
GPT as a judge & 2025 & Annotation judgement with symbolic representations. \\
Audio Flamingo 3 & 2025 & General audio-language modelling and long audio. \\
Music Flamingo & 2025 & Specialized music audio-language modelling. \\
HumMusQA & 2026 & Expert-written musical reasoning questions. \\
GaMMA & 2026 & Global/temporal questions and segmentation evaluation. \\
MusTBench & 2026 & Temporal-grounding questions across five tasks. \\
MUB (this draft) & 2026 & Identity-masked global labels and full-track intervals; validation incomplete. \\
\bottomrule
\end{tabularx}
\caption{Positioning relative to verified related work. Model papers and benchmark papers serve different roles and are not compared as equivalent datasets.}
\label{tab:related}
\end{table}

\section{Dataset Construction}
\label{sec:dataset}
\subsection{Source and Label Provenance}
The source is an existing annotation archive from a music-annotation workflow described as professional annotation. Records contain timestamped structural intervals and per-interval tempo, tonality, time signature, and additional attributes. MUB uses only the attributes relevant to the four tasks. It does not independently verify annotator qualifications or listening procedures.

The upstream testset contains 2,345 task records. Its builder takes the last available annotation submission and derives modal global labels from segment attributes. A unique plurality is not necessarily a strict majority, and consistency across segments is not agreement across independent annotators. The current MUB selector does not enforce a completed quality-review status. Among the 500 selected entries, 399 have one recorded operator and 117 have one recorded submission. Revision counts must therefore not be interpreted as independent expert verification.

\subsection{Eligibility and Deduplication}
\label{sec:selection}
Candidates must have all three global probes, at least five structural intervals, and duration between 90 and 600 seconds. Exact title--artist string pairs are deduplicated before identities are removed, keeping the task with the most segments. This yields 1,111 candidate pairs.

This process removes repeated metadata pairs but does not establish unique compositions or unique audio: alternate titles, recordings, and versions may remain. It also does not use a recording-rights or licence field. The selected 500 recordings cannot currently be described as the lowest-copyright-risk subset.

\subsection{Sampling}
Strata combine four tempo bins (below 70, 70--99, 100--129, and at least 130 BPM) with key/mode. With random seed 42, the selector processes smaller strata first, shuffles within each, and uses a remaining-budget allocation with a minimum of one candidate while budget remains. A final random fill draws from the unused candidates. This is not proportional sampling. Time signature, duration, genre, and structural complexity are not sampling strata.

\subsection{Identity Masking and Audio Matching}
\label{sec:anonymization}
Each entry receives an opaque identifier from MUB-00001 to MUB-00500. Table~\ref{tab:anonymization} distinguishes evaluator-side records from model inputs. Gold labels are available to the evaluator and must never be sent with an inference request.

\begin{table}[tbp]
\centering\small
\begin{tabularx}{\linewidth}{@{}p{0.40\linewidth}XX@{}}
\toprule
Field & Evaluator records & Model request \\
\midrule
Original title and artist & Removed & Excluded \\
Original platform song/task IDs & Removed & Excluded \\
Account and operator names & Removed & Excluded \\
Opaque MUB identifier & Retained & Routing only \\
Partial-file fingerprint & Retained & Must be excluded \\
Gold labels & Retained & Excluded \\
Audio and task instruction & Matched locally & Required \\
\bottomrule
\end{tabularx}
\caption{Intended separation of scoring data and inference inputs. The generated JSONL files contain routing fields; they are not themselves a ready-made model-visible message.}
\label{tab:anonymization}
\end{table}

The current \texttt{audio\_hash} is SHA-256 of the first 1,048,576 bytes of the compressed source file, truncated to 16 hexadecimal characters (64 bits). It is a partial-file fingerprint, not a full-file checksum, perceptual hash, or anonymity guarantee. Re-encoding may change it, and matching against a known catalogue could identify the recording. It must remain evaluator-side. The prompt records currently include this field for local matching, so an inference adapter must forward only the audio and task text. Embedded audio tags and filenames also require sanitization; the current selection script does not certify that this has occurred.

Audio is not distributed. Supplying an accessible recording with a different encoding is insufficient for byte-prefix matching, and the DSP code does not retrieve or download missing recordings. A controlled private mapping and a legitimate audio-access route are needed for reproducible external evaluation.

\subsection{Collection Statistics}
All 24 major/minor key combinations occur, with 268 major and 232 minor labels. Table~\ref{tab:stats} gives the main statistics. The time-signature distribution is highly imbalanced: 492 recordings are 4/4, four are 3/4, and four are 6/8.

\begin{table}[tbp]
\centering\small
\begin{tabularx}{\linewidth}{@{}Xr@{}}
\toprule
Property & Value \\
\midrule
Recordings & 500 \\
BPM range & 47--180 \\
BPM: 25th percentile / median / 75th percentile & 75 / 101 / 128 \\
BPM below 70 / 70--99 / 100--129 / at least 130 & 87 / 156 / 150 / 107 \\
Major / minor & 268 / 232 \\
Key/mode combinations & 24 \\
Time signatures: 4/4 / 3/4 / 6/8 & 492 / 4 / 4 \\
Duration range / median (seconds) & 92--535 / 228 \\
Intervals per recording: range / median & 5--22 / 9 \\
Total structural intervals & 4,676 \\
Distinct lowercased structural labels & 34 \\
\bottomrule
\end{tabularx}
\caption{Statistics computed from the current benchmark manifest.}
\label{tab:stats}
\end{table}

The most frequent lowercased labels are chorus (1,231), verse (1,114), intro (460), outro (427), pre-chorus (392), interlude (324), bridge (282), and post-chorus (154). The manifest retains free-text and capitalization variation; lowercasing is used here for descriptive counts, not as an implemented label-scoring procedure.

\section{Tasks and Inference Protocol}
\label{sec:tasks}
\paragraph{Tempo.}
Predict a scalar tempo in beats per minute. The reference is the upstream modal segment tempo, not a continuous tempo curve or a fresh consensus tapping experiment. Half/double-tempo ambiguity and tempo changes can make a single global label incomplete. The current prompt asks for an integer between 40 and 220.

\paragraph{Key.}
Predict tonic pitch class and major/minor mode. References are parsed from modal segment tonalities, with enharmonic pitch names mapped to pitch classes. Modulation, ambiguous tonal centres, and other modes are outside the current representation. Relative major and minor are distinct tonal interpretations even when they share a key signature.

\paragraph{Time signature.}
Predict a fraction such as 4/4, 3/4, or 6/8. The reference is a modal segment label, so changing metre is reduced to a single answer. The dominance of 4/4 limits conclusions drawn from aggregate accuracy.

\paragraph{Structural segmentation.}
Return intervals with start and end times in seconds and a short section label. The current scorer measures interval overlap only; it does not score the correctness of labels such as verse or chorus. Full-track segmentation requires localization across a recording but is not, under this metric alone, a demonstration of semantic formal analysis.

\subsection{External Assistance and Model Inputs}
\label{sec:tool-constraints}
The generated prompts ask the model to rely solely on the provided audio and explicitly prohibit web search, internet access, downloads, code execution, external tools, external DSP, fingerprinting services, music recognition, and database lookup. They also discourage use of prior recording-specific knowledge.

For the intended native-audio evaluation track, the deployment must disable network/search, downloads, code execution, external DSP services, and model-callable tools. This is a protocol requirement; no enforcing inference harness is currently supplied. Prompt text cannot guarantee compliance, and an instruction not to use prior knowledge cannot remove memorized recording information from model parameters. A tool-assisted track, if later added, must be reported separately.

An eventual evaluation must record model identifier and version, inference date, sampling settings, audio decoding and resampling, context limits, truncation or chunking, output parser, retries, refusal rates, tool configuration, and verified input fields. The same audio coverage must be used across models. Full-recording tasks cannot be compared fairly when one model receives only an undisclosed excerpt.

\section{Metrics and Implementation Audit}
\label{sec:metrics}
\subsection{Tempo and Time Signature}
Tempo is a hit when
\begin{equation}
 |\widehat b-b|\leq 0.04b,
\end{equation}
where $b$ is the reference BPM. The 4\% value is a benchmark convention, not a claim of perceptual indistinguishability. Median relative error is also reported. The current metric does not credit octave-related tempos. Time signature is matched after extracting and normalizing its numerator and denominator.

\subsection{Legacy Key Metric}
\label{sec:key-metric}
The current implementation accepts exact tonic/mode matches, or a different-mode prediction whose tonic plus three semitones equals the reference tonic modulo twelve. This second condition is erroneous and directional: it accepts A minor against C major, rejects the reverse, and can accept C major against E-flat minor. Accordingly, the stored 25.4\% DSP and 9.2\% constant-prior results are labelled \emph{legacy key scores}, not conventional key accuracies.

A corrected protocol should use exact tonic/mode accuracy as the primary measure and, if desired, a separately named symmetric relative-key measure. For example, map a minor tonic $p$ to $(p+3)\bmod 12$ and leave a major tonic unchanged, then compare the resulting relative-major tonics. Re-scoring the same saved predictions with that symmetric rule gives 29.4\% for DSP and 11.0\% for the constant prior; these are diagnostic recalculations, not outputs of a repaired released scorer. Exact-key and relative-key results must not be mixed. A uniform predictor over 24 keys has expected accuracy 1/24 for exact matching and 2/24 for exact-or-relative matching.

\subsection{Interval Segmentation}
For each reference interval, the scorer chooses the highest-IoU unmatched prediction and counts a match at IoU at least 0.5. It then computes precision, recall, and F1 from the number of matches; reported mean F1 averages per-recording scores. This is greedy matching and may depend on interval order. A small number of intervals does not establish equivalence to optimal assignment.

Section labels are ignored. Thus, this is an interval-overlap score, not an independent boundary-tolerance score or a label-accuracy score. The current parser does not comprehensively validate coverage, continuity, bounds, or malformed intervals. A stronger protocol should include explicit interval validation, boundary metrics, label-aware measures, and a documented matching algorithm.

\subsection{Prediction Coverage}
The current evaluator iterates over submitted predictions: unknown identifiers are skipped, absent recordings are excluded from denominators, and repeated identifiers can be counted repeatedly. Null fields in an otherwise present row count as misses. This is unsuitable for unrestricted benchmark submissions. A corrected version should require exactly one prediction per expected identifier and score failures against a fixed 500-recording denominator. The two stored baseline files each contain 500 unique, matching identifiers, so this defect does not alter their reported aggregate results.

\section{Reference Baselines and Local Results}
\label{sec:baselines}
\subsection{Waveform DSP Reference}
Audio is decoded with FFmpeg to 22,050 Hz mono PCM. The implementation uses a 2,048-sample Fourier window and a 512-sample hop. Tempo is estimated from spectral-flux autocorrelation over a 60--180 BPM search range. Key is estimated by correlating aggregate pitch-class energy with major/minor key profiles, following the pitch-profile approach associated with Krumhansl~\cite{krumhansl}. The time-signature heuristic compares periodicities at two, three, and four beat multiples and always uses denominator four; it cannot output 6/8.

No structural segmentation is implemented in this baseline: every structural output is null. Its zero F1 therefore measures abstention, not the limits of DSP segmentation. These heuristics are reference systems rather than state-of-the-art baselines, theoretical lower bounds, or ceilings on achievable performance. Agreement with them cannot independently validate annotation provenance.

\subsection{Answer-Distribution Reference}
A constant predictor uses the benchmark's own median tempo (101 BPM), modal key (C major), and modal metre (4/4), with no intervals. Because these constants are derived from test references, this is a descriptive test-distribution prior, not a trained text-only model or an estimate of out-of-sample generalization.

\subsection{Results and Evidence}
Table~\ref{tab:baseline-results} reports scores from existing local prediction and report files. An arithmetic audit compared all saved predictions with the manifest and reproduced 268 tempo hits, 127 legacy key hits, and 183 metre hits for DSP. Both files have complete 500-recording identifier coverage. This confirms consistency of the saved artifacts under the current scoring rules; it does not authenticate the execution history. Audio DSP was not rerun for this manuscript revision.

\begin{table}[tbp]
\centering\small
\begin{tabularx}{\linewidth}{@{}Xrr@{}}
\toprule
Measure & DSP & Constant prior \\
\midrule
BPM hit rate ($\pm$4\%) & 53.6\% & 9.0\% \\
BPM median relative error & 2.9\% & 22.3\% \\
Legacy key score (defective rule) & 25.4\% & 9.2\% \\
Time-signature accuracy & 36.6\% & 98.4\% \\
Mean interval F1 (no intervals predicted) & 0.000 & 0.000 \\
\bottomrule
\end{tabularx}
\caption{Local reference results under the current implementation. Legacy key scores are not standard exact-key accuracies. Both systems abstain on segmentation. No MLLM results are included.}
\label{tab:baseline-results}
\end{table}

The DSP tempo estimator exceeds the constant prior, but this alone does not identify why individual predictions fail or establish label correctness. The constant metre predictor's 98.4\% result shows that aggregate metre accuracy is dominated by the class distribution. A model can use informative audio and still underperform a strong prior on this imbalanced task. Likewise, a positive interval F1 can arise from generic segmentation and does not prove semantic understanding.

The result artifacts currently reside in a locally ignored results directory and are not tracked by the repository. A public reproducible release must package the prediction files, software versions, commands, and data/metric version identifiers. The current local repository should not be described as an already published benchmark release.

\section{Limitations, Rights, and Release Conditions}
\label{sec:limitations}
\paragraph{Rights are not established by anonymization.}
Withholding audio reduces redistribution exposure but does not establish rights to use recordings or redistribute annotations. No recording-level licence verification, copyright-risk ranking, or annotation-rights gate was applied. The previous README's annotation-licence declaration was not evidence of authorization and has been withdrawn from the current documentation. A public release needs documented rights for its actual contents and an access policy for recordings. This draft does not claim a rights-cleared or lowest-risk 500-recording subset.

\paragraph{Annotation reliability remains unmeasured.}
The workflow provenance does not replace an independent annotation study. Global labels come from a last-submission plurality across sections; submission counts do not establish inter-annotator agreement. Tempo parsing, tonal normalization, structural ambiguity, review status, and alternative valid analyses require auditing. A stratified double-annotation sample with adjudication is needed before stronger gold-quality claims can be made.

\paragraph{Coverage and imbalance.}
The representation restricts tonality to major/minor and globalizes potentially changing properties. Only eight recordings have a non-4/4 reference, so balanced accuracy or macro-averaged scores should supplement overall metre accuracy, with uncertainty clearly reported. Genre diversity has not been quantified, and metadata-string deduplication does not certify 500 distinct compositions.

\paragraph{Leakage and access.}
Removing explicit metadata does not prevent recognition of familiar audio, memorization, or training contamination. A byte-prefix fingerprint may also permit linkage. Model-visible hashes, audio tags, and revealing filenames must be excluded. Without distributable audio or a controlled access route, independent reproduction remains difficult.

\paragraph{Incomplete empirical validation.}
There are no MLLM runs, confidence intervals, audio-ablation controls, or implemented segmentation baselines in this version. The key metric and coverage handling require versioned fixes before new model comparisons. A future experiment should compare audio-present and audio-absent conditions, log tool restrictions, validate outputs, and report failures rather than silently excluding them. These are requirements for further validation, not results obtained here.

\section{Conclusion}
MUB assembles 500 identity-masked recordings with tempo, major/minor key, time signature, and structural-interval references from an existing annotation workflow. Recent work, especially GaMMA, already covers overlapping global and temporal tasks; MUB's present value lies in its specific resource, input-masking protocol, and transparent implementation audit. Saved baseline predictions expose the importance of answer priors and metric definitions, but do not establish contemporary model capabilities. Rights documentation, annotation validation, corrected versioned scoring, and controlled MLLM experiments are required before presenting it as a validated public benchmark.

\end{document}